\documentclass[12pt,preprint]{aastex}

\shorttitle{Evolution of the most luminous quasar}
\shortauthors{Wang et al.}

\begin{document}


\title{Probing the interstellar medium and star formation of the Most Luminous Quasar at z=6.3}


\author{Ran Wang\altaffilmark{1},
Xue-Bing Wu\altaffilmark{1,2}, 
Roberto Neri\altaffilmark{3}, 
Xiaohui Fan\altaffilmark{4,1},
Fabian Walter\altaffilmark{5},
Chris L. Carilli\altaffilmark{6},
Emmanuel Momjian\altaffilmark{6},
Frank Bertoldi\altaffilmark{7},
Michael A. Strauss\altaffilmark{8},
Qiong Li\altaffilmark{2},
Feige Wang\altaffilmark{2},
Dominik A. Riechers\altaffilmark{9},
Linhua Jiang\altaffilmark{1},
Alain Omont\altaffilmark{10},
Jeff Wagg\altaffilmark{11},
Pierre Cox\altaffilmark{12}}
\altaffiltext{1}{Kavli Institute of Astronomy and Astrophysics at Peking University, 
No.5 Yiheyuan Road, Haidian District, Beijing, 100871, China}
\altaffiltext{2}{Department of Astronomy, School of Physics, Peking University, No. 5 Yiheyuan Road, Haidian District, Beijing, 100871, China}
\altaffiltext{3}{Institute de Radioastronomie Millimetrique, St. Martin d'Heres, F-38406, France}
\altaffiltext{4}{Steward Observatory, University of Arizona, 933 N Cherry Ave., Tucson, AZ, 85721, USA}
\altaffiltext{5}{Max-Planck-Institute for Astronomy, K$\rm \ddot o$nigsstuhl 17, 69117 Heidelberg, Germany}
\altaffiltext{6}{National Radio Astronomy Observatory, PO Box 0, Socorro, NM, USA 87801}
\altaffiltext{7}{Argelander-Institut f$\rm \ddot u$r Astronomie, University of Bonn, Auf dem H$\rm \ddot u$gel 71, 53121 Bonn, Germany}
\altaffiltext{8}{Department of Astrophysical Sciences, Princeton University, Princeton, NJ, USA, 08544}
\altaffiltext{9}{Astronomy Department, Cornell University, 220 Space Sciences Building, Ithaca, NY 14853, USA}
\altaffiltext{10}{Institut d'Astrophysique de Paris, UMR 7095, CNRS and Universite Pierre et Marie Curie, Paris, France}
\altaffiltext{11}{Square Kilometre Array Organisation, Jodrell Bank Observatory, Jodrell Bank, Macclesfield SK11 9DL}
\altaffiltext{12}{Joint ALMA Observatory, Alonso de C\'{o}rdova 3107, Vitacura, Santi-ago, Chile}

\begin{abstract}

We report new IRAM/PdBI, JCMT/SCUBA-2, and VLA observations of the ultraluminous quasar 
SDSSJ010013.02+280225.8 (hereafter, J0100+2802) at z=6.3, which hosts the most massive   
supermassive black hole (SMBH) of $\mathrm{1.24\times10^{10}\,M_{\odot}}$ known at z$>$6. 
We detect the [C II] 158$\mu$m fine structure line and 
molecular CO(6-5) line and continuum emission at 353 GHz, 260 GHz, and 3 GHz 
from this quasar. The CO(2-1) line and the underlying 
continuum at 32 GHz are also marginally detected. The [C II] and CO detections suggest 
active star formation and highly excited molecular gas in the quasar host 
galaxy. The redshift determined with the [C II] and CO 
lines shows a velocity offset of $\rm \sim1000\,km\,s^{-1}$ from that 
measured with the quasar Mg II line. 
The CO (2-1) line luminosity provides direct constraint on the molecular gas mass 
which is about $\rm (1.0\pm0.3)\times10^{10}\,M_{\odot}$. We estimate the 
FIR luminosity to be $\rm (3.5\pm0.7)\times10^{12}\,L_{\odot}$, and the UV-to-FIR 
spectral energy distribution of J0100+2802 is consistent with the templates 
of the local optically luminous quasars. The derived [C II]-to-FIR 
luminosity ratio of J0100+2802 is 0.0010$\pm$0.0002, which is slightly higher than the values 
of the most FIR luminous quasars at z$\sim$6. We investigate the constraint on 
the host galaxy dynamical mass of J0100+2802 based on the [C II] line spectrum. 
It is likely that this ultraluminous quasar lies above the local 
SMBH-galaxy mass relationship, unless we are viewing the system at a 
small inclination angle.
% and the host galaxy is still forming stars at a rate 
%of $\mathrm{\lesssim600\,M_{\odot}\,yr^{-1}}$. 
\end{abstract}





\section{Introduction}

An ultraluminous quasar, SDSS J010013.02+280225.8 (hereafter, J0100+2802)  
with a bolometric luminosity of $\mathrm{L_{bol}=4.3\times10^{14}\,L_{\odot}}$ and a 
supermassive black hole (SMBH) mass of $\mathrm{M_{BH}\approx1.2\times10^{10}\,M_{\odot}}$, 
was discovered at z=6.3 \citep{wu15}. 
This is by far the most optically luminous object, with the most massive SMBH,  
among the $\sim$50 quasars known at z$>$6 \citep{fan06,willott10,jiang15,venemans15,
banados14,matsuoka16}, and the SMBH mass is also comparable to
the highest value found in the local universe \citep{mcconnell11,thomas16}. It demonstrates that such a 
rare, $\mathrm{10^{10}\,M_{\odot}}$ SMBH can be formed as early 
as z$\gtrsim$6.3, when the age of the universe was only 890 Myr. 

Recent submillimeter and millimeter [(sub)mm] surveys detected strong 
far-infrared (FIR) continuum, molecular CO, and [C II] 158$\mu$m 
fine structure line emission in the host galaxies of a number of 
quasars at z$>$5.7 (e.g., \citealp{bertoldi03,priddey03,robson04,
walter04,carilli07,wang10,wang13,omont13,willott15,venemans16}). The results argue for 
significant host galaxy evolution with active star formation in these young quasars 
in the early universe, and the [C II] 
line images at sub-second resolution constrain radius of the nuclear starburst region to  
be a few kpc \citep{leipski13,wang13,venemans16}. The gas dynamics measured 
with the CO and [C II] observations suggest SMBH-host mass ratios a factor of a few 
to one order of magnitude higher than the present-day 
value \citep{walter04,wang13,willott15,venemans16}.

The most massive z$>$6 quasar-galaxy systems studied in previous millimeter and radio 
observations are objects with SMBH masses of a few $\mathrm{10^{9}\,M_{\odot}}$ and
$\mathrm{L_{bol}\lesssim10^{14}\,L_{\odot}}$ (e.g., \citealp{wang13,willott15,venemans16}). 
The discovery of J0100+2802 provides a unique chance to study the 
quasar-galaxy co-evolution in the system that hosts the most massive known SMBH 
at the earliest epoch. In this work, we report our new observations of the 
(sub)mm and radio continuum, [C II] 158$\mu$m, and molecular CO lines 
from the host galaxy of J0100+2802. The observations are described in Section 2, 
the results are presented in Section 3. We discuss the quasar 
spectral energy distribution (SED), FIR continuum, [C II], and CO line ratios, 
and constraints of the host galaxy dynamical mass in Section 4, and summarize 
the results in Section 5.
A $\Lambda$-CDM cosmology with $\mathrm{H_{0}=71km\,s^{-1}\,Mpc^{-1}}$, 
$\mathrm{\Omega_{M}=0.27}$ and $\mathrm{\Omega_{\Lambda}=0.73}$ is adopted 
throughout this paper \citep{spergel07}.

\section{Observations}

\subsection{PdBI observations}

We observed the [C II] 158$\mu$m ($\mathrm{^{2}P_{3/2}-^{2}P_{1/2}}$)  
and molecular CO(6-5) line emission from J0100+2802 using the Plateau de 
Bure interferometer (PdBI). 
The observations were carried out in Summer 2014 in D-configuration using five antennas. 
We used the WideX wide-band correlator in dual polarization with a total bandwidth
of 3.6GHz. We set up the correlator centered at the frequency of 260.35 GHz for the [C II] line and 94.72 GHz 
for the CO(6-5) line. This corresponds to the quasar Mg II line redshift of z=6.30 \citep{wu15}. 
The flux density scale was determined based on observations of the 
standard flux density calibrator MWC349, and the typical calibration uncertainties are 10\% 
at 3 mm and 20\% at 1 mm. The phase was checked about every 22min by observing a nearby 
phase calibrator, J0112+321. We reduced the data using the Grenoble Image and Line 
Data Analysis System (GILDAS) software \citep{guilloteau00}. The maps were made using 
natral weighting. The FWHM synthesized beam sizes are $\mathrm{2.00''\times1.66''}$ 
and $\mathrm{5.41''\times4.31''}$ at the [C II]
and CO(6-5) frequencies, respectively. The final rms sensitivity of 
the [C II] line observation is 2.3 $\mathrm{mJy\,beam^{-1}}$ per 46 $\mathrm{km\,s^{-1}}$ 
channel with a total on-source integration time of 4.5 hours, and the rms of the CO(6-5) 
observation is 0.42 $\mathrm{mJy\,beam^{-1}}$ per 63 $\mathrm{km\,s^{-1}}$ channel in 13.2 hours.

\subsection{VLA observations}

We observed the CO(2-1) line emission with the Karl G. Jansky Very Large Array (VLA) 
in Ka-Band ($\rm \sim32$ GHz) in C-configuration in Oct-Nov 2014, and the 3 GHz 
radio continuum in S-band in A-configuration in June 2015. The observations were performed using the WIDAR correlator with 
the 8-bit sampler in dual polarization. The setup covers a total bandwidth of 2 GHz with sixteen 128 MHz 
spectral windows and the channel width in each spectral window is 2 MHz. 
We have spent a total on-source observing time of 7.2 hours for the CO(2-1) 
line observation and 2.1 hours for the 3 GHz continuum. The flux density calibration was carried out using the standard VLA 
calibrator 3C48 with typical calibration uncertainties better than 5\%. The phases were checked with nearby point sources. 

We reduced the data using the Common Astronomy Software Applications package (CASA v4.4) 
and the VLA calibration pipeline version 1.3.4. 
The FWHM synthesized beam size is $\mathrm{0.74''\times0.68''}$ 
at 32 GHz and $\mathrm{0.65''\times0.54''}$ at 3 GHz, using robust weighting.
For the CO(2-1) line observations, we binned the data 
to a channel width of 57 $\mathrm{km\,s^{-1}}$, and obtained a typical 1$\sigma$ rms noise of 
0.067 $\mathrm{mJy\,beam^{-1}}$ per channel. The channels affected by radio frequency 
interference were removed in the S-Band observations, resulting in a useful frequency 
range of $\sim$2.4 to 4 GHz and bandwidth of about 1.6 GHz. 
The final continuum sensitivity at 3 GHz was 3.1 $\mathrm{\mu Jy\,beam^{-1}}$. 

\subsection{SCUBA-2 observations}

We observed the 450$\mu$m (666 GHz) and 850$\mu$m (353 GHz) dust continuum from J0100+2802 using the 
Submillimetre Common-User Bolometer Array 2 (SCUBA-2, \citealp{holland13}) on the James Clerk 
Maxwell Telescope (JCMT) in Nov 2015. 
The observations were carried out in Band 2 weather 
conditions (i.e. $\mathrm{0.05<\tau_{225GHz}<0.008}$), and in 'CV DAISY' mode which is designed for point/compact source observations. 
The beam size of SCUBA-2 is 7.9$''$ at 666 GHz and 13$''$ at 353 GHz. 
We observed the target in four 30min scans with a total on-source time of $\sim$2 hours. The data was 
reduced using the STARLINK SCUBA-2 pipeline for faint point sources \citep{chapin13}, and we 
obtained a 1$\sigma$ point-source sensitivity of 1.2 mJy 
at 353 GHz and 10 mJy at 666 GHz.

\section{Results}

We summarize the measurements of redshift, line flux, FWHM line width, and luminosity of the [C II], CO(6-5), and CO(2-1) 
lines, as well as the continuum in Table 1. 
The [C II] and CO(6-5) lines are clearly detected (Figure 1), as well as the continuum emission at 353 GHz, 260 GHz, and 3 GHz. 
We fit a Gaussian line profile to the [C II] and CO(6-5) line spectra to determine 
the redshifts and line widths, and integrate the continuum-subtracted data over the line-emitting channels to get the line flux. 
The redshifts and line widths measured on the [C II] and CO(6-5) lines are consistent with each other within the errors, and we 
adopt $\mathrm{z_{[CII]}=6.3258\pm0.0010}$ as the redshift of the quasar host galaxy. 
We obtain tentative detections for the CO(2-1) line and the 32 GHz continuum. 
The central frequency and width of the CO(2-1) line are difficult to constrain due
to the poor signal to noise ratio (SNR) of the spectrum.
We integrate the visibility data over the velocity range of the CO(6-5) line emission (i.e.,
800 $\mathrm{km\,s^{-1}}$ to 1300 $\mathrm{km\,s^{-1}}$ in Figure 1), and the intensity map
shows a 3.6$\sigma$ peak about 0.27$''$ away from the optical quasar position.
Considering the measurement uncertainties and the astrometric mismatch between radio 
and optical frame, this is consistent
with the quasar optical position. We also measure the 32 GHz continuum emission to 
be $\mathrm{S_{32GHz}=14.8\pm4.3\,\mu Jy}$ by averaging all the line-free spectral windows. 
We subtract the continuum from the line-emitting
channels, resulting in a CO(2-1) line flux of $\mathrm{0.038\pm0.013\,Jy\,km\,s^{-1}}$ at the peak position.  
Due to the poor SNR of the intensity map, we cannot obtain reliable
measurement of the source size, or address if there are more extended components with
lower surface brightness. 

The quasar Mg II line emission from J0100+2802 measures a redshift 
of $\rm z_{MgII}=6.301\pm0.006$ (Figure 3 in\citealp{wu15}), 
which is blueshifted by $\mathrm{1020\pm250\,km\,s^{-1}}$ compared 
to $\mathrm{z_{[CII]}}$. Similar blueshifted Mg II lines with velocity 
offsets of a few hundred to 1700 $\mathrm{km\,s^{-1}}$ are also detected in 
several other z$>$6 quasars \citep{willott15,venemans16}. 
Such large Mg II line velocity offsets with respect to the quasar 
host galaxy redshift are rare at low redshift \citep{richards02}, and may suggest outflowing gas in the broad 
line region at an early evolutionary phase. A larger sample of high-z quasars 
with $\mathrm{z_{MgII}}$ and $\mathrm{z_{[CII]}}$ measurements is required to investigate the relationship 
between MgII velocity offset and quasar luminosities \citep{venemans16}
and address the origin of the blushifted quasar Mg II line.

\section{Discussion}

\subsection{Continuum emission and spectral energy distribution}

The new observations we present here measure the rest-frame FIR to 
radio continuum emission from J0100+2802. 
We plot the SED of J0100+2802 in Figure 2 \citep{wu15}, together with the data from SDSS, 
the Two Micron All Sky Survey (2MASS), and the Wide-field 
Infrared Survey Explorer (WISE). The templates of optically 
luminous quasars from \citet{elvis94} and \citet{richards06} are also plotted for 
comparison. We scale the templates to the 2MASS H-band flux, and the (sub)mm  
measurements show no excess compared to the FIR emission of the scaled templates. 
We cannot rule out that the FIR continuum emission from J0100+2802 
is dominated by thermal emission from the AGN-heated dust. However, the detections 
of [C II] and CO line emission do argue for star forming interstellar medium (ISM) 
in the nuclear region (see Section 4.2). 
We fit the (sub)mm flux densities to an optically thin 
graybody model with a dust temperature of $\mathrm{T_{dust}=47}$ K and an emissivity 
index of $\mathrm{\beta=1.6}$ from the (sub)mm-detected quasars at z$>$1.7 \citep{beelen06}. 
This results in a FIR (42.5$\mu$m-122.5$\mu$m) luminosity of 
$\mathrm{L_{FIR}=(3.5\pm0.7)\times10^{12}\,L_{\odot}}$, and a 8$\mu$m-1000$\mu$m IR luminosity of 
$\mathrm{L_{IR}=(5.0\pm0.9)\times10^{12}\,L_{\odot}}$. 
The star formation rate (SFR) estimated with the IR luminosity using the relation in \citet{kennicutt98} 
is $\mathrm{850\,M_{\odot}\,yr^{-1}}$. As the fraction contributed from the torus is unknown, 
this constrains the upper 
limits of the star formation-powered FIR emission and SFR in the host galaxy. 
The dust mass calculated based on the above dust temperature and emissivity index 
assumption is $\mathrm{M_{dust}=(2.0\pm0.4)\times10^{8}\,M_{\odot}}$. We adopt a dust absorption 
coefficient of $\mathrm{\kappa_{d}(\lambda)\sim\lambda^{-\beta}}$ and 
$\mathrm{\kappa_{d}(125\mu m)=18.75\,cm^{2}\,g^{-1}}$ \citep{hildebrand83} in the caluclation. 
If a higher dust temperature of 60 K is adopted (i.e., the highest temperature seen among the
submillimeter/millimeter-detected z$\sim$6 quasars, \citealp{leipski13}), the FIR luminosity and SFR increase
by a factor of 1.7, and the dust mass will decrease by a factor of 1.8. If we adopt the lowest dust
temperature of 40 K found by \citet{leipski13}, the FIR and SFR with decrease by a
factor of 1.5, and the dust mass increases by a factor of 1.6 (Table 1).

The VLA Ka and S-band observations measure the rest-frame 234 GHz and 22 GHz
continuum emission, respectively. The corresponding continuum luminosities 
are $\rm \nu L_{\nu,234GHz}=(5.8\pm1.7)\times10^{8}\,L_{\odot}$ 
and $\rm \nu L_{\nu,22GHz}=(3.9\pm0.1)\times10^{8}\,L_{\odot}$. 
We calculate the radio loudness of this object adopting the definition 
of $R=f_{\nu,5GHz}/f_{\nu,4400\AA}$ \citep{kellermann89}. The rest-frame 4400$\AA$ 
flux density ($f_{\nu,4400\AA}$) is interpolated using the scaled Richards et al. (2006) optically 
luminous quasar template (Figure 1), and the rest-frame 5 GHz flux density ($f_{\nu,5GHz}$) 
is extrapolated from 22 GHz data assuming a power-law radio continuum of 
$\rm f_{\nu}\propto\nu^{\alpha}$. This constrains the radio loudness to be $\rm R=0.9$ 
for a steep spectrum of $\rm \alpha=-0.9$, or $\rm R=0.2$ for a flat spectrum of $\rm \alpha=-0.06$ 
(see below for the discussion of spectral index), indicating that this object 
is radio quiet. However, the 234 GHz and 22 GHz luminosities are more than an 
order of magnitude higher than the thermal bremsstrahlung or nonthermal synchrotron continuum 
expected from star forming activities (e.g., Condon 1992; Yun et al. 2000; Yun \& Carilli 2002; 
Zakamska et al. 2016). For instance, the thermal bremsstrahlung or nonthermal synchrotron 
continuum estimated with a SFR of $\rm 850\,M_{\odot}\,yr^{-1}$ using the formulae 
in Yun \& Carilli (2002) contributes $<$5\% of the detected continuum emission at both 
frequencies. Additionally, the thermal dust emission has only a minor contribution 
of 12\% to 21\% to the 234 GHz flux density 
based on the graybody models with $\rm T_{dust}=60\,K$ to 40 K described above. 
Thus, the continuum emission in J0100+2802 at frequencies of 234 GHz and lower 
is dominated by the radio activity of the central AGN.  

In Figure 1, we adopt the graybody model with $\mathrm{T_{dust}=47\,K}$ to remove the dust continuum 
at 234 GHz and fit a power-law ($\mathrm{f_{\nu}\sim\nu^{\alpha}}$) to the remaining 
234 GHz flux density and the 22 GHz data. 
This estimates the 234-to-22 GHz 
two-point spectral index to be $\mathrm{\alpha^{234GHz}_{22GHz}=-0.90\pm0.15}$. 
We also measure the spectral index within the S-band by 
averaging the data in every two 128MHz spectral windows from the observed 
frequencies of 2.4 GHz to 4 GHz. In contrast to the steep spectrum indicate by 
$\rm \alpha^{234GHz}_{22GHz}$, the best result fit to the 
seven data points in S-band yield a flat spectrum with 
$\mathrm{\alpha_{22GHz}=-0.06\pm0.22}$. J0100+2802 provides a great example 
to study the radio activity in the most optically luminous and radio quiet nucleus 
at the earliest epoch. We will need further observations at multiple radio 
frequencies to check if the radio continuum is indeed 
flattened or inverse around 22 GHz. Additionally, milli-arcsecond resolution observations 
through Very Long Baseline Interferometry are needed to measure the spatial 
extent of the radio source in this object and address if there are multiple components contribute to the 
detected radio continuum as was widely discussed in the radio quiet quasars 
at lower redshift (e.g., Ulvestad et al. 1999, 2005)

\subsection{Luminosity ratios and gas masses}

Based on the FIR luminosity drived in the previous section, 
we constrain the [C II]-to-FIR luminosity ratio to be $\rm 0.0010\pm0.0002$.
This is comparable to the values found in other [C II]-detected z$>$5.7 quasars that have similar moderate FIR luminosities 
of $\mathrm{L_{FIR}\lesssim1\times10^{12}\,L_{\odot}}$ to $\mathrm{\sim4\times10^{12}\,L_{\odot}}$,
and higher than that from the more FIR luminous 
objects ($\rm L_{FIR}\geq5\times10^{12}\,L_{\odot}$, \citealp{wang13}), i.e., following the trend of decreasing FIR
luminosity with increasing [C II]-to-FIR luminosity ratios defined by the
high-redshift quasars and starburst galaxies \citep{maiolino09,riechers13,debreuck14,willott15,
gullberg15,venemans16,munoz15,narayanan16}. 
The SFR estimated with the [C II] luminosity is $\mathrm{560\,M_{\odot}\,yr^{-1}}$, 
adopting the SFR-[C II] luminosity relation for high-redshift galaxies 
in \citet{delooze14}. This is consistent with the SFR estimates based-on the dust continuum. 

The CO observations measure a CO (6-5)-to-(2-1) line flux 
ratio of $\mathrm{8.4\pm3.6}$ and a line luminosity ratio 
of $\mathrm{{L'}_{CO(6-5)}/{L'}_{CO(2-1)}=0.94\pm0.40}$. 
This is consistent with the ratios found 
in other CO-detected z$>$5.7 quasars within the errors \citep{wang11,stefan15}, suggesting 
that the detected CO emisison is likely from a highly-excited 
molecular gas component peaked at J=6 or higher \citep{riechers09,gallerani14,stefan15}. 
If we assume that the low-J CO transitions are thermalized, i.e., $\mathrm{{L'}_{CO(1-0)}={L'}_{CO(2-1)}}$ 
(e.g., \citealp{carilli13}), and adopt a luminosity-to-mass conversion factor of 
$\rm \alpha_{CO}=0.8\,M_{\odot}\,/K\,km\,s^{-1}\,pc^{2}$ from the local 
ultraluminous infrared galaxies \citep{downes98}, the detected CO(2-1) line 
flux yields a molecular gas mass of $\mathrm{M_{mol}=(1.0\pm0.3)\times10^{10}\,M_{\odot}}$. 
However, the total molecular gas mass in the quasar host could be larger considering that some of the CO (2-1) line flux
from the low surface brightness region and the line wings might be missing
due to the poor SNR of the observation.

We compare the luminosity ratios of the FIR continuum, [C II], and CO lines from J0100+2802 to the models 
of Photo dissociation region (PDR\footnote{The line and FIR ratio models are taken from the Photo Dissociation Region Toolbox (http://dustem.astro.umd.edu/pdrt/).}, \citealp{kaufman99,kaufman06,pound08}) in Figure 3, to investigate the gas density n 
and the incident far-ultraviolet radiation field $\rm G_{0}$ (in units of the Habing Field, 
$\rm 1.6\times10^{-3}\,ergs\,cm^{-2}\,s^{-1}$) of the interstellar medium in the quasar host galaxy. 
As a one-side illuminated slab geometry was adopted in the model, we here devide the 
optically thin FIR\footnote{The FIR luminosity is re-calculated in the range 
of 30$\mu$m to 1000$\mu$m to match the definition in the models.} and [C II] line 
emission by a factor of 2 to match the condition in the model. 
We also multiply $\rm L_{[C II]}$ by a factor of 0.7 to estimate and exclude the [C II] emission 
from the diffuse region \citep{stacey91,colbert99}. 
According to Figure 3, the overlap region of the three luminosity ratios 
suggests $\rm G_{0}$ of a few thousand and n$\rm \gtrsim1\times10^{5}\,cm^{-3}$. 
These are only preliminary constraints on the physical condition of the star 
forming region in the host galaxy of J0100+2802. There are still large uncertainties 
in the calculation of FIR luminosity. The fractions of [C II] and CO emission 
from the PDR region are also not well determined. A larger FIR luminosity with a higher 
$\rm T_{dust}$ will result in a higher $\rm G_{0}$, and any contribution from 
the X-ray dominated regions (XDR) to the detected CO (6-5) line 
emission (e.g., \citealp{gallerani14}) will result in a higher $\rm L_{[C II]}/L_{CO(6-5)}$ 
and a lower $\rm L_{CO(6-5)}/L_{CO(2-1)}$ from the PDR region and move the best-fit gas 
density to $\rm n<10^{5}\,cm^{-3}$. 

We estimate the atomic gas mass within the PDR region from 
the [C II] line luminosity using Equation (1) in \citet{haileydunsheath10},  
adopting the best fit parameters indicated in Figure 3 (i.e., $\rm G_{0}\sim4000$, 
$\rm n\sim10^{5.1}\,cm^{-3}$, and a corresponding PDR surface temperature 
of $\mathrm{T\sim550\,K}$, \citealp{kaufman99}), a C$^+$ abundance 
of $\mathrm{1.4\times10^{-4}}$, and a critical density 
of $\mathrm{n_{crit}=2.7\times10^{3}\,cm^{-3}}$.   
The derived atomic gas mass in the PdR region 
is $\rm M_{atomic,PDR}\sim3\times10^{9}\,M_{\odot}$, which is about 30\% of the molecular 
gas mass estimated with the CO (2-1) line.  
The mass ratio associated with the detected CO, [C II], and FIR dust emission 
is  $(M_{mol}+M_{atomic,PDR})/M_{dust}\sim65$. 
This is comparable to the mass ratios   
of other CO-detected quasars and dusty 
starburst galaxies at high-z \citep{wang10,michalowski10,riechers13}. 

\subsection{Constraint on host galaxy dynamical mass}

Among the quasars known at z$>$6, J0100+2802 has the most massive SMBH with a mass of 
$\mathrm{(1.24\pm0.19)\times10^{10}\,M_{\odot}}$ \citep{wu15,derosa11}. We use the [C II] 
line width ($\mathrm{FWHM_{[C II]}}$) to present a preliminary estimate of the 
host galaxy mass for this object. We assume that the line is 
from a rotating gas disk and the circular velocity can be estimated 
as $\mathrm{v_{cir}=0.75FWHM_{[C II]}/sin(i)}$, where i is the 
inclination angle between the rotation axis of the disk and 
the line of sight. The [C II] line emission from J0100+2802 is unresolved in 
our PdBI observation at 2$''$ resolution. 
According to the recent [C II] observations of other z$\sim$6 quasars at sub-arcsecond 
resolution, the typical FWHM major axis sizes of the [C II] emission are 
about 2$\sim$4 kpc \citep{wang13,willott13,willott15,venemans16}. 
Thus, we here estimate the FWHM [C II] source size for J0100+2802 to be $\mathrm{3\pm1}$ 
kpc (see also \citealp{willott15,venemans16}) and set the diameter of the gas disk D as   
1.5$\times$ the FWHM [C II] source size, i.e., $\mathrm{D=4.5\pm1.5}$ kpc.
The host galaxy dynamical mass is then 
$\rm M_{dyn}(M_{\odot})=1.16\times10^{5}{v_{cir}}^{2}D=[2.6(D/4.5\,kpc)\pm1.6]\times10^{10}/sin^{2}(i)\,M_{\odot}$. 
The error inlcudes the unceratainties from both the line width 
and the assumed disk size. 

We plot $\mathrm{M_{BH}}$ vs. $\mathrm{M_{dyn}}$ for J0100+2802 and 
other z$>$5.7 quasars in Figure 4 \citep{wang13,willott13,willott15,banados15,venemans16}, 
comparing to the SMBH-to-bulge mass relation of local galaxies from \citet{kormendy13}, i.e., 
$\rm M_{BH}/(10^{9}\,M_{\odot})=0.49(M_{bulge}/10^{11}\,M_{\odot})^{1.16}$. 
For J0100+2802 as well as other z$>$5.7 quasars that have SMBH mass measurements based on the 
quasar Mg II line emisison \citep{derosa11,derosa14,willott13,willott15}, 
we follow \citet{willott15} and add a 0.3 dex uncertainty to the error bar of the SMBH 
mass to account for the scatter of the calibration \citep{shen08}. 
For the sample from \citet{wang13} which do not have 
SMBH mass measurements, we adopt 
the relationship between the 1450$\AA$ luminosity and the quasar bolometric 
luminosity from \citet{venemans16}, and calculate the Eddington luminosities and 
SMBH masses assuming a typical Eddington ratio 
and a scatter of $\mathrm{log(L_{bol}/L_{Edd})=-0.3\pm0.3}$ from \citet{derosa11}. 
The $\mathrm{M_{dyn}}$ for most of the z$>$5.7 quasars are estimated based
on [C II] observations \citep{wang13,willott13,willott15,banados15,venemans16}. 
The only exception is the z=6.42 quasar SDSS J114816.64+525150.3, 
in which the [C II]-emitting gas at $>$1.5 kpc scale is turbulent \citep{cicone15} 
and the CO size is used \citep{riechers09,stefan15} in the $\mathrm{M_{dyn}}$ 
calculation. According to Figure 4, for any inclination angle of $\mathrm{i\geq10^{\circ}}$, 
J0100+2802 is above the local $\mathrm{M_{BH}-M_{bulge}}$ relation and 
the $\pm$0.3 dex area of the intrinsic scatter (i.e. the gray area in Figure 3). 
As was discussed in \citet{willott15}, most of the z$\sim$6 quasars with SMBH masses on 
order of $\mathrm{10^{8}\,M_{\odot}}$ are close to the trend of local galaxies,  
while the more luminous and massive objects tend to be above this 
trend (see also \citealp{venemans16}). This suggests that the SMBH grows faster 
than the quasar host galaxies in these most massive systems at 
the earliest epoch, unless all these $\mathrm{M_{BH}>10^{9}\,M_{\odot}}$ 
quasars are close to face-on. However, as there is no resolved image for J0100+2802 
yet, we does not rule out the possibility that the gas is unvirialized and the [C II] 
line width cannot probe the disk circular velocity. 

\section{Summary}

We detected [C II], CO, and (sub)mm and radio continuum emission 
in the host galaxy of the quasar J0100+2802 which hosts the most massive SMBH known at z$\geq$6. 
The detections probe the properties of the young quasar host at an early evolutionary stage: 
The (sub)mm continuum indicates moderate FIR emission and constrains the
SFR to be $\mathrm{\leq850\,M_{\odot}\,yr^{-1}}$. The CO and [C II] lines  
estimate the gas mass and gas-to-dust mass ratio that are within the range of other millimeter-detected 
quasars at z$\sim$6. The [C II]-to-FIR luminosity ratio J0100+2802 is higher than
that of the most FIR luminous quasars at z$>$5.7, i.e., following the
trend of increasing $\mathrm{L_{[C II]}/L_{FIR}}$ with decreasing $\mathrm{L_{FIR}}$
found with the high-z quasars and star forming systems. 
The quasar Mg II line emission detected in previous near-infrared 
spectroscopic observation \citep{wu15} is blueshifted by about 1000 $\rm km\,s^{-1}$ 
compared to the host galaxy redshift measured by the [C II] and CO lines.
The host dynamical mass estimated with the [C II] line width suggest 
that the SMBH is likely to be overmassive, compared to the local 
relation, though furhter constraints on the gas kinematics and disk inclination angle are still required.

\acknowledgments

The data presented in this paper are based on observations
under project number S14CY with the IRAM Plateau de Bure Interferometer,
projects 14B151 and 15A494 with the VLA, and project M15BI055
with JCMT/SCUBA-2. IRAM is supported by INSU/CNRS (France),
MPG (Germany) and IGN (Spain). The National Radio Astronomy Observatory
is a facility of the National Science Foundation operated under
cooperative agreement by Associated Universities, Inc.
The James Clerk Maxwell Telescope is operated by the East Asian
Observatory on behalf of The National Astronomical Observatory of
Japan, Academia Sinica Institute of Astronomy and Astrophysics, the
Korea Astronomy and Space Science Institute, the National Astronomical
Observatories of China and the Chinese Academy of Sciences (Grant No. XDB09000000),
with additional funding support from the Science and Technology
Facilities Council of the United Kingdom and participating universities
in the United Kingdom and Canada.
We are thankful for the supports from the National
Science Foundation of China (NSFC) grants No.11373008, 11533001, 
the Strategic Priority Research Program $"$The Emergence of Cosmological Structures$"$
of the Chinese Academy of Sciences, grant No. XDB09000000,
the National Key Basic Research Program of China 2014CB845700, 
the Ministry of Science and Technology of China under grant 2016YFA0400703.
RW acknowledge supports from the Thousand Youth
Talents Program of China, the NSFC grants No. 11443002 and 11473004.
XF acknowledge supports
from NSF Grants AST 11-07682 and 15-15115. 
We thank M. Wolfire for providing line ratio maps used in the PDRToolBox.

%\email{rwangkiaa@pku.edu.cn}.

{\it Facilities:} \facility{IRAM (PdBI)}, \facility{VLA}, \facility{JCMT (SCUBA-2)}.

\clearpage

%% the file, uncomment the \plotone, \plottwo, and
%% If you need a layout that cannot be achieved with \plotone or
%% \plottwo, you can invoke the graphicx package directly with the
%% \includegraphics command or use \plotfiddle. For more information,
%% please see the tutorial on "Using Electronic Art with AASTeX" in the
%% documentation section at the AASTeX Web site,
%% http://www.journals.uchicago.edu/AAS/AASTeX.
%%
%% The examples below also include sample markup for submission of
%% supplemental electronic materials. As always, be sure to check
%% the instructions to authors for the journal you are submitting to
%% for specific submissions guidelines as they vary from

%% This example uses \plotone to include an EPS file scaled to
%% 80% of its natural size with \epsscale. Its caption
%% has been written to indicate that additional figure parts will be
%% available in the electronic journal.
\begin{table}
\caption{Line and continuum properties of J0100+2802}
\begin{tabular}{lc}
\hline \noalign{\smallskip}
\hline \noalign{\smallskip}
[C II] redshift & 6.3258$\pm$0.0010 \\
\multicolumn{1}{l}{[C II] FWHM ($km\,s^{-1}$)}  & 300$\pm$77 \\
\multicolumn{1}{l}{[C II] line flux ($Jy\,km\,s^{-1}$)} & 3.36$\pm$0.46 \\
\multicolumn{2}{l}{[C II] line luminosity:} \\
\multicolumn{1}{c}{($L_{\odot}$)} & $(3.56\pm0.49)\times10^{9}$ \\
\multicolumn{1}{c}{($K\,km\,s^{-1}\,pc^{2}$)} & $(1.62\pm0.22)\times10^{10}$ \\ 
CO(6-5) redshift & 6.3264$\pm$0.0024 \\
CO(6-5) FWHM ($km\,s^{-1}$) & 498$\pm$225 \\
CO(6-5) line flux ($Jy\,km\,s^{-1}$) & 0.32$\pm$0.084 \\
\multicolumn{2}{l}{CO(6-5) line luminosity:} \\
\multicolumn{1}{c}{($L_{\odot}$)} & $(1.23\pm0.32)\times10^{8}$ \\
\multicolumn{1}{c}{($K\,km\,s^{-1}\,pc^{2}$)} & $(1.17\pm0.30)\times10^{10}$ \\
CO(2-1) line flux ($Jy\,km\,s^{-1}$) & 0.038$\pm$0.013 \\
\multicolumn{2}{l}{CO(2-1) line luminosity:} \\
\multicolumn{1}{c}{($L_{\odot}$)} & $(4.89\pm1.67)\times10^{6}$ \\
\multicolumn{1}{c}{($K\,km\,s^{-1}\,pc^{2}$)} & $(1.25\pm0.43)\times10^{10}$ \\
\multicolumn{2}{l}{CO(1-0) line luminosity [derived from the CO (2-1) line]:} \\
\multicolumn{1}{c}{($L_{\odot}$)} & $(6.11\pm2.09)\times10^{5}$ \\
\multicolumn{1}{c}{($K\,km\,s^{-1}\,pc^{2}$)} & $(1.25\pm0.43)\times10^{10}$ \\
666 GHz continuum (mJy) & $<$30 \\
353 GHz continuum (mJy) & 4.1$\pm$1.2 \\
260 GHz continuum (mJy) & 1.35$\pm$0.25 \\
94.5 GHz continuum (mJy) & $<$0.1 \\
32 GHz continuum ($\mu$Jy) & 14.8$\pm$4.3 \\
3 GHz continuum ($\mu$Jy) & 104.5$\pm$3.1 \\ 
\multicolumn{2}{l}{FIR luminosity ($L_{\odot}$):} \\
\multicolumn{1}{c}{$T_{dust}=47$ K, $\beta =1.6$ } & $(3.5\pm0.7)\times10^{12}$ \\
\multicolumn{1}{c}{$T_{dust}=40$ K, $\beta =1.6$ } & $(2.4\pm0.5)\times10^{12}$ \\
\multicolumn{1}{c}{$T_{dust}=60$ K, $\beta =1.6$} & $(6.0\pm1.2)\times10^{12}$ \\
\multicolumn{2}{l}{dust mass ($M_{\odot}$):} \\
\multicolumn{1}{c}{$T_{dust}=47$ K, $\beta =1.6$} & $(2.0\pm0.4)\times10^{8}$ \\
\multicolumn{1}{c}{$T_{dust}=40$ K, $\beta =1.6$} & $(3.1\pm0.6)\times10^{8}$ \\
\multicolumn{1}{c}{$T_{dust}=60$ K, $\beta =1.6$} & $(1.1\pm0.2)\times10^{8}$ \\
\noalign{\smallskip} \hline
\end{tabular}
\end{table}

\begin{figure}
\epsscale{.80}
\plotone{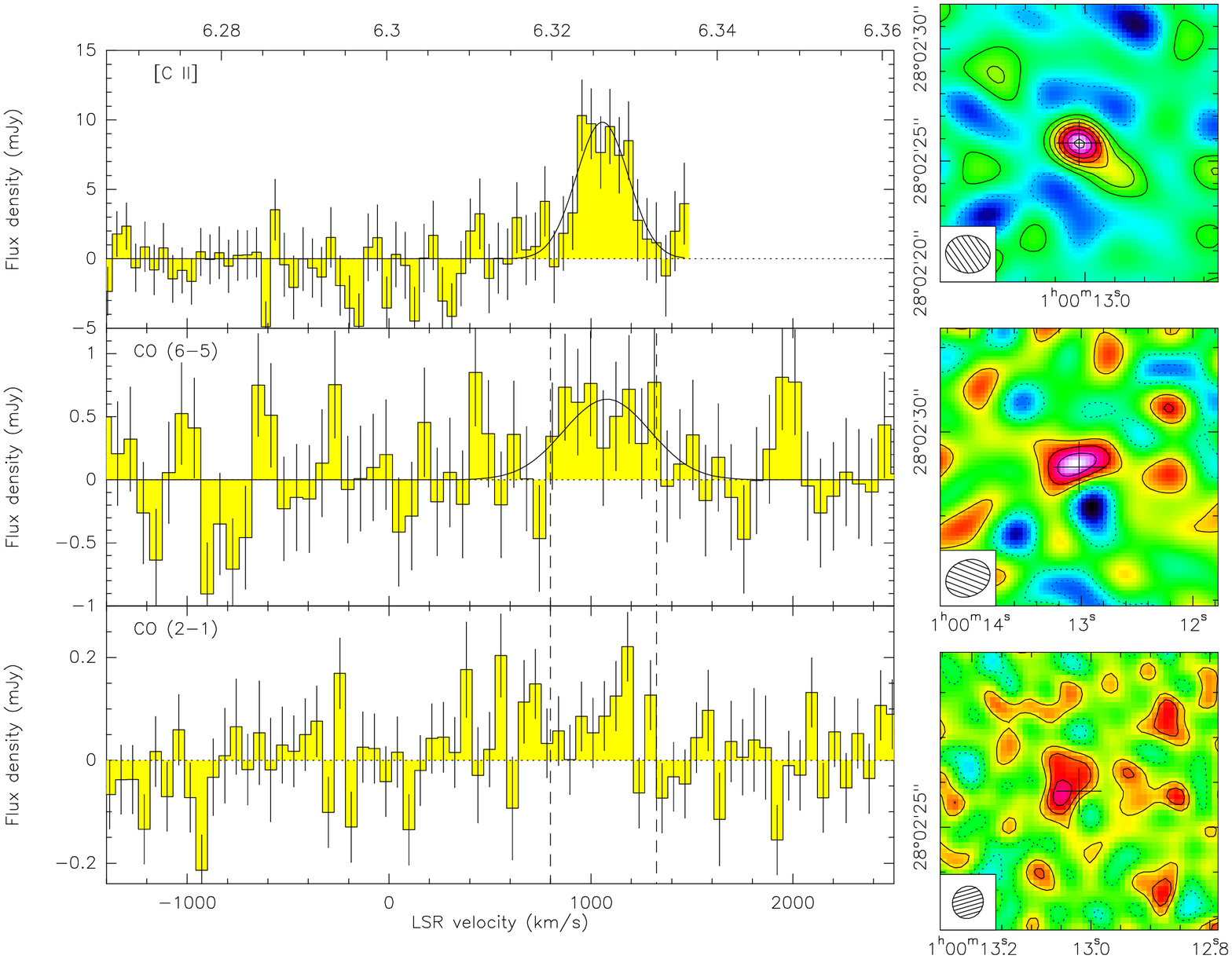}
\caption{The line spectra (left) and velocity  
integrated map (right) of the [C II] 158$\mu$m (upper panel), CO(6-5) (middle panel), and 
CO(2-1) line emission (lower panel) from J0100+2802. The bottom 
abscissa shows the radio velocity in Kinematic Local Standard of Rest (LSRK) frame, where zero velocity 
corresponds to $\mathrm{z_{MgII}=6.30}$ \citep{wu15}. The top abscissa 
denotes the corresponding redshift. The contours for 
the [C II] line intensity map are [-2, -1, 1, 2, 3, 4, 5, 6, 7]
$\times0.77\,mJy\,beam^{-1}$, and [-2, -1, 1, 2, 3]
$\times0.15\,mJy\,beam^{-1}$ for the CO (6-5) line. 
The CO(2-1) line intensity map is obtained by 
integrating over the velocity range defined by the CO(6-5) 
line (vertical dashed lines), and contours 
are [-2, -1, 1, 2, 3]$\times23\,\mu Jy\,beam^{-1}$. The cross in each panel 
denotes the position of the optical quasar, and the synthesized 
beams are plotted at the bottom left of each map. \label{fig1}}
\end{figure}

%% Here we use \plottwo to present two versions of the same figure,
%% one in black and white for print the other in RGB color
%% for online presentation. Note that the caption indicates
%% that a color version of the figure will be available online.
%%

\begin{figure}
%\plottwo{f2.eps}{f2_color.eps}
\epsscale{.80}
\plotone{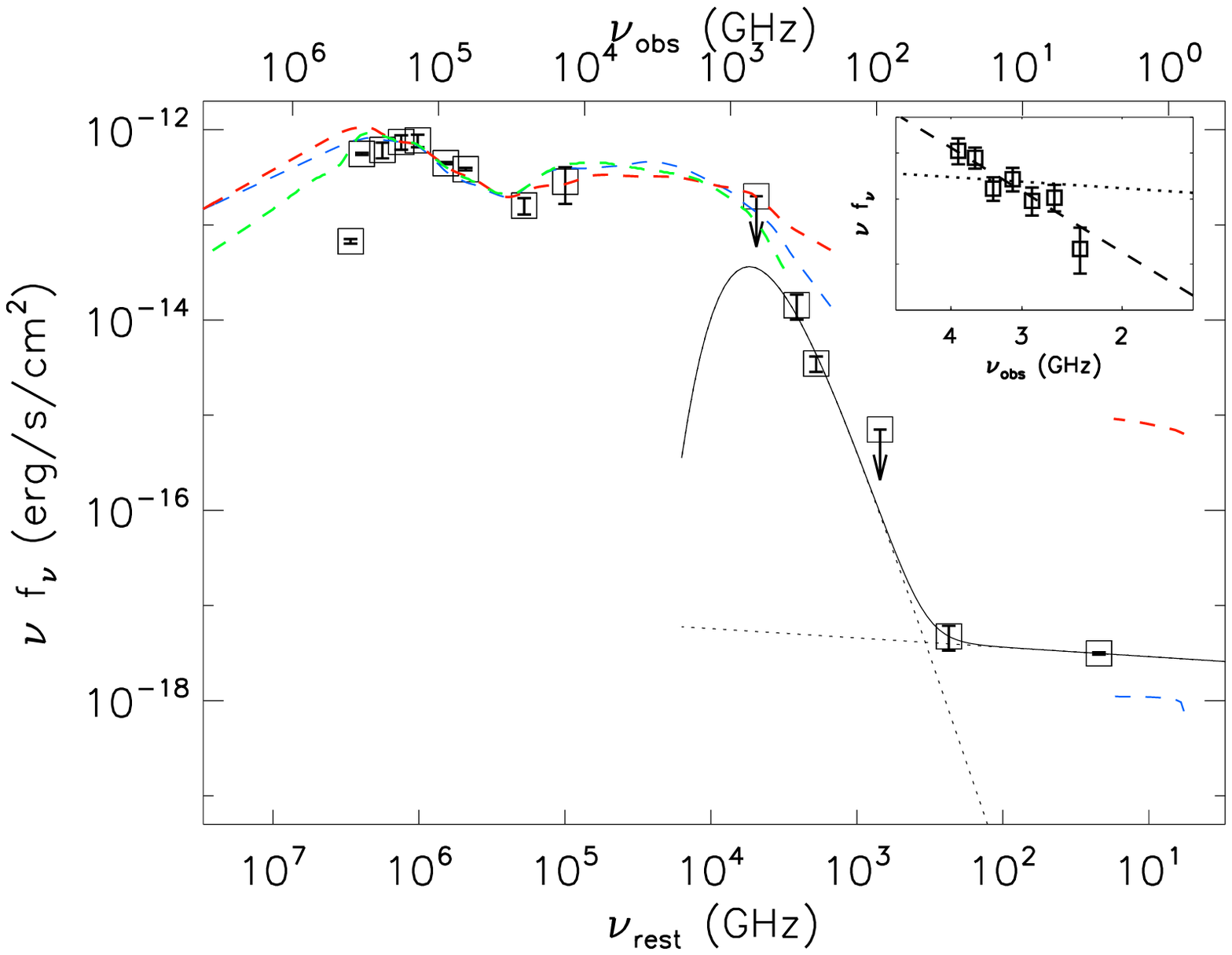}
\caption{The rest-frame UV-to-radio SED of J0100+2802. The squares show the data from SDSS, 2MASS, 
and WISE \citep{wu15}, and the new measurements in this work. 
The arrows denote $\rm 3\sigma$ upper limits. The templates of 
local radio quiet (blue line) and radio loud 
quasars (red line, \citealp{elvis94}) and SDSS optically luminous 
quasars (green line, \citealp{richards06}) are plotted and scaled to the 2MASS H-band flux. 
The dotted lines are the graybody and power-law (with $\mathrm{f_{\nu}\sim\nu^{\alpha}}$ and $\mathrm{\alpha^{234GHz}_{22GHz}=-0.90}$) 
fit to the FIR and radio continuum, respectively, and the black solid 
line shows the total emission from these two components. 
The inner panel shows the S-band continuum fluxes measured 
in every 258 MHz frequency bins from 2.3 to 4 GHz. 
The dotted line is the power-law with $\mathrm{\alpha^{234GHz}_{22GHz}=-0.90}$ fitted to the Ka-band continuum and the S-band flux density 
averaged over the total 1.6 GHz bandwidth, while the dashed line is the best fitting result to 
only the S-band data with $\mathrm{\alpha_{22GHz}=-0.06}$.} 
\end{figure}

%% This figure uses \includegraphics to scale and rotate the still frame

%\begin{figure}
%\epsscale{.80}
%\plotone{c2co.eps}
%\includegraphics[angle=90,scale=.50]{f3.eps}
%\caption{The [C II] vs. CO (6-5) line luminosities of the 
%seven [C II] and CO (6-5)-detected quasars at z$\sim$6 (Bertoldi et al. 2003b;  
%Carilli et al. 2007; Wang et al. 2010, 2011a). 
%The red star shows the luminosities of SDSS J0100+2802 reported in this work. 
%A linear fitting to the data yields a relation of 
%$\rm log (L_{[C II]}/L_{\odot}) = (2.66\pm1.20)+(0.83\pm0.15)\times log (L_{CO(6-5)/L_{\odot}})$ 
%(solid line), while the dashed line shows the fitting result when we fix the slope to 1.
%\label{fig3}}
%\end{figure}

\begin{figure}
\epsscale{.80}
\plotone{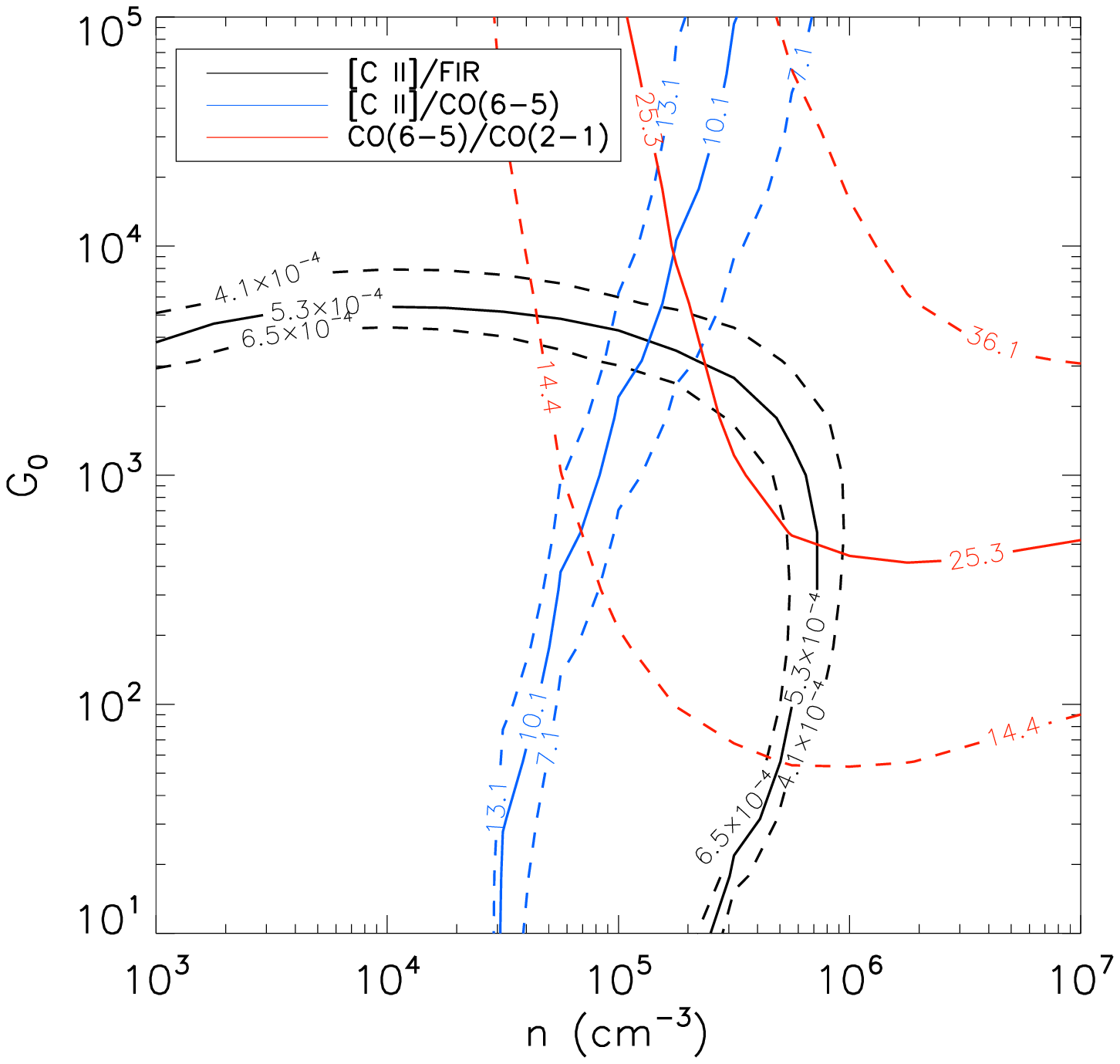}
\caption{Luminosity ratios of $\rm L_{[C II]}/L_{FIR}$ (black), 
$\rm L_{[C II]}/L_{CO(6-5)}$ (blue), and $\rm L_{CO(6-5)}/L_{CO(2-1)}$ (red) 
compared to the PDR models of Kaufman et al. (1999) in different conditions 
of radiation field $\rm G_{0}$ and gas density n. Luminosities are all in unit of $\rm L_{\odot}$. 
The dashed contours denote 
the $\rm \pm1\sigma$ error bars from the observing uncertainties. 
We adopt the optically thin graybody model with $\rm T_{dust}=47$ K and $\rm \beta=1.6$ 
to calculate $\rm L_{FIR}$. 
The $L_{FIR}$ here is integrated in the wavelength range of 30$\mu$m to 1000$\mu$m 
to match the FIR definition in the model. We also assume that 70\% of the 
detecte [C II] line emission is from the PDR region. Thus the $\rm L_{[C II]}/L_{FIR}$ value in the plot 
is lower than the value of 0.001 quoted in Section 4.1.
We also devide $\rm L_{[C II]}$ and $\rm L_{FIR}$ by a factor of 2 to match the 
one-side illuminated geometry in the model. }
\end{figure}

\begin{figure}
\epsscale{.80}
\plotone{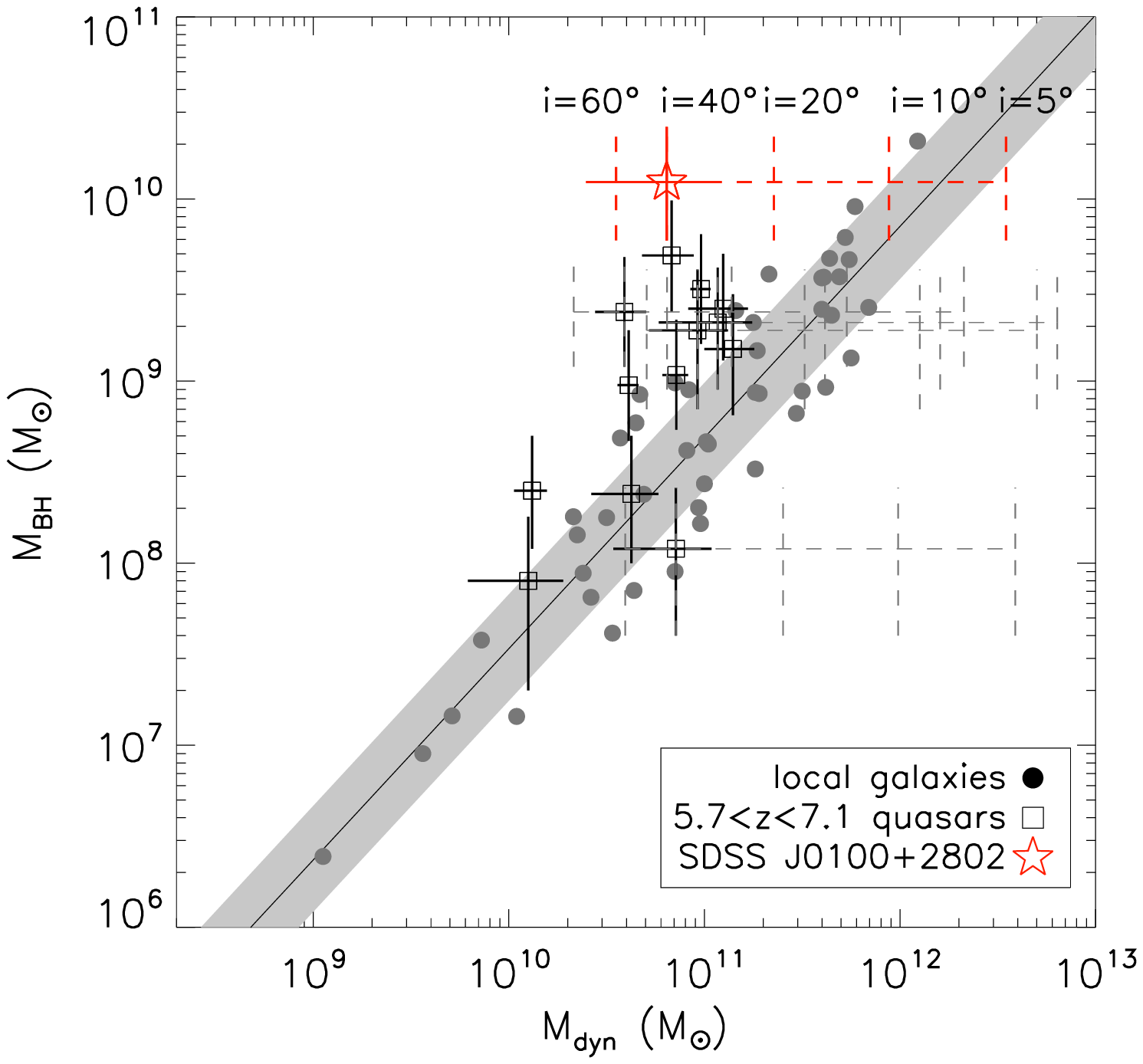}
\caption{$\mathrm{M_{BH}}$ vs. $\mathrm{M_{dyn}}$ of the [C II]-detected z$>$5.7 quasars. 
The $\mathrm{M_{dyn}}$ for z$>$5.7 quasars are estimated based 
on [C II] observations, except one object, SDSS 
J114816.64+525150.3 at z=6.42, in which the [C II]-emitting gas at $>$1.5 
kpc scale is turbulent and the CO size is  
adopted \citep{riechers09,wang13,willott13,willott15,cicone15,venemans16}. 
The red star shows J0100+2802 in this work. 
For objects that do not have an inclination angle estimated with the 
resolved [C II] image, we show $\mathrm{M_{dyn}}$ calculated with 
different inclination angles (dashed lines). The solid line and 
the gray region show the local relationship with $\pm0.3$ dex intrinsic scatter. 
The gray circles are the sample of local galaxies \citep{kormendy13}.
}
\end{figure}

\end{document}